\documentclass{iopart}   
\usepackage{iopams}  
\usepackage{epsf}  
\usepackage{times}  
 
\eqnobysec

\begin{document}

\jl{1}

\paper{A Note on a hyper-cubic Mahler measure and associated Bessel integral.}

\author{ML Glasser\dag}

\address{\dag\ Department of Physics, Clarkson University, Potsdam\\ NY 13699-5820, USA}
\vskip .3in
\centerline{\it Dedicated to Fa Yueh Wu in recognition of his 80-th birthday}

\begin{abstract}
The Mahler measure for the n-variable polynomial $k+\sum(x_j+1/x_j)$ is reduced to a single integral of the $n-th$ power of the modified Bessel Function $I_0$. Several special cases are examined in detail.
\end{abstract}

\pacs{02.30.Gp, 05.50.+q}

\submitted

\section{Introduction}
Interest in the Mahler measure
$$m(P)=\int_0^1dt_1\cdots\int_0^1dt_n\ln|P(e^{2\pi it_1},\cdots, e^{2\pi it_n})|\eqno(1)$$
of a polynomial $P(x_1,\cdots, x_n)$ has grown since Deninger[1] and Boyd[2] formulated a number of conjectures connecting them with the $L-$functions of an important class of elliptic  curves. More recently, Bertin[3] and Rogers[4,5] have extended this connection to hypergeometric identities and multi-dimensional lattice sums. For restricted ranges of the coefficients of certain $P$, $m(P)$ reduces to a variety of entropic integrals of interest in statistical physics, such as  the spanning tree generating functions, studied extensively by Wu and his co-workers[6,7,8,9], as well as their  derivatives, the corresponding lattice Green functions[10].

The aim of this note is to examine the Mahler measure of  the hyper-cubic  ``polynomial"
$$k+\sum_{j=1}^n(x_j+1/x_j).\eqno(2)$$
With $k=2z$, $z\ge n$, after a simple change of integration variables,  (1) becomes
$$m(P)=\ln 2+J_n(z)$$
$$J_n(z)=\frac{1}{\pi^n}\int_0^{\pi}dx_1\cdots\int_0^{\pi}dx_n\ln[z\pm \sum_{j=1}^n\cos x_j].\eqno(3)$$
(The sign is irrelevant). In 1987 the case $J_3(3)$ was first studied by A. Rosengren[11] and subsequently Joyce and Zucker[12] investigated $J_d(d)$ by reducing it to the single integral 
$$J_d(d)=\int_0^{\infty}[e^{-t}-e^{-td}I_0^d(t)]\frac{dt}{t}.\eqno(4)$$
By means of an asymptotic expansion Joyce and Zucker could approximate the integral (4) and obtain $J_d(d)$ to 50 place accuracy for $d=2,\cdots, 10.$ In addition they worked out 15 terms of the asymptotic expansion for $d\rightarrow\infty$.  The principal result of this note is the derivation of an expression  for $J_n(z)$ similar to (4) for arbitrary $z>n$ and to explore some consequences.

\section{Calculation}
    
    We start with the integral representation
    $$\ln A=-\gamma-A\int_0^{\infty}dx \, \ln x\, e^{-Ax}.\eqno(5)$$
    Then, with $A=z+\sum\cos x_j$, the familiar representation
    $$I_0(t)=\frac{1}{\pi}\int_0^{\pi}dx \, e^{\pm t \cos x}\eqno(6)$$
    and the fact that $I_0'(t)=-I_1(t)$, we have
    $$J_n(z)=-\gamma-\int_0^{\infty}dx\,\ln x\, e^{-zx}I_0^{n-1}(x)[zI_0(x)+nI_1(x)].\eqno(7)$$
    Being careful about the cancellation of the divergences at the lower limit, the second integral can be integrated by parts and recombined with the first to yield
    $$J_n(z)=-\gamma+\int_0^{\infty}\frac{dx}{x}[\Theta(1-x)-e^{-zx}I_0^n(x)]=\int_0^{\infty}\frac{dx}{x}[e^{-x}-e^{-zx}I_0^n(x)],\eqno(8)$$
    where $\Theta$ denotes the unit step function. Equation (8) is valid for $z\ge n\ge0$ and is consistent with Joyce and Zucker's expression (4) since
  $$  \int_0^1\frac{1-e^{-x}}{x}dx-\int_1^{\infty}\frac{e^{-x}}{x}dx=\gamma.\eqno(9)$$

\section{Results and conclusions}

By comparing (8) with the known values of $J_n(z)$ for $n=1,2$ we obtain several, apparently, new Bessel integrals:
$$\int_0^{\infty}\frac{dx}{x}[\Theta(x-1)-e^{-zx}I_0(x)]=\gamma+\ln\left(\frac{z+\sqrt{z^2-1}}{2}\right)\eqno(10)$$
$$\int_0^{\infty}\frac{dx}{x}[e^{-x}-e^{-zx}I_0^2(x)]=\ln(z)-\frac{1}{2z^2}\;_4F_3(1,1,\frac{3}{2},\frac{3}{2};2,2,2;\frac{4}{z^2})\eqno(11)$$
$$\int_0^{\infty}dx\,\ln x\, e^{-zx}[zI_0^2(x)+2I_0(x)I_1(x)]=$$
$$\ln(1/z)-\gamma+\frac{1}{2z^2}\;_4F_3(1,1,\frac{3}{2},\frac{3}{2};2,2,2;\frac{4}{z^2}).\eqno(12)$$
Note that
$$\;_4F_3(1,1,\frac{3}{2},\frac{3}{2};2,2,2;z)=$$
$$\frac{16}{\pi z}\int_0^{\sqrt{z}}\frac{du}{u}[{\bf K}(u)-{\bf K}(0)].\eqno(13)$$
Also $\;_4F_3(1,1,3/2,3/2;2,2,2;1)=16(\pi\ln 2-2{\bf G})/\pi$.

Next, from the techniques introduced in Ref.[13] one finds
$$\int_0^{\infty}e^{-zt}I_0^3(t)dt=$$
$$\frac{3(r+2z)}{10(z^2-9)}F\left[-\frac{16(2z(r-z)+9}{(z^2-9)^2}\right]-\frac{(z^2+15)r-2z^3+18z}{5((z^2-9)(z^2+3)}F\left[-\frac{32z(r-z)+144}{(z(r-z)+3)^4}\right]$$
$$+\frac{18(r-2z)}{5(z^2-9)(z^2+3)}\eqno(14)$$
where $r=\sqrt{z^2-9}$ and $F[\xi]=\;_3F_2(1/4,1/2,3/4;1,1;\xi)$.  By integrating (14) with respect $z$ from $z=5$, where the second term of (14) vanishes and the first term simplifies considerably, to $z$ (this was carried out using Mathematica), we arrive at the closed-form expression

$$J_3(z)=\int_0^{\infty}\frac{dx}{x}[e^{-x}-e^{-zx}I_0^3(x)]=$$
$$-\frac{1}{5}\{\ln\left[\frac{2R_0}{(z^2+3)^3}\right] +\frac{9}{8}\frac{R_1^3}{R_2^4}\;_5F_4\left(1,1,
5/4,3/2,7/4;
2,2,2,2;
16\frac{R_1^3}{R_2^4}\right)$$
$$+\frac{3}{8}\frac{R_1}{R_3^4}\;_5F_4\left(1,1,
5/4,3/2,7/4;
2,2,2,2
;16\frac{R_1}{R_3^4}\right)\},\eqno(15)$$
where
$$R_0=z^3-5z-(z^2-1)\sqrt{z^2-3}$$
$$R_1=2z^2-9-2z\sqrt{z^2-9}$$
$$R_2=z^2-9-z\sqrt{z^2-9}$$
$$R_3=z^2-3-z\sqrt{z^2-9}\eqno(15)$$
For $3\le z\le5$ the use of (15) requires the analytic continuation of the  hypergeometric function which is examined in Appendix A; for $z>5$ (15) is correct as stated. An equivalent expression appears in the work of Guttmann and Rogers[16]. The value (15) is  a companion to Joyce's expressions[15] for the FCC lattice.

\vskip .1in
Can one proceed in this way? In view of the work by Glasser and Guttmann [17] concerning
$$\int_0^{\infty}e^{-zx}I_0^4(x)dx
\eqno(16)$$
which includes its differential equation and series expansions,
$n=3$ is probably as far as one can go in terms of known hypergeometric functions defined by a single series.

\ack 
The author thanks Tony Guttmann, John Zucker and Mathew Rogers for substantial help with this project.



\vskip1cm

\appendix
\section{ Analytic continuation of $\;_5F_4(1,1,5/4,3/2,7/4;2,2,2,2;z)$}

By examining the respective series, it is clear that
$$\;_5F_4(1,1,5/4,3/2,7/4;2,2,2,2;z)=\frac{1}{z}\int_0^zdt\;_4F_3(1,5/4,3/2,7/4;2,2,2;t).\eqno(A1)$$
Similarly,
$$\;_4F_3(1,5/4,3/2,7/4;2,2,2;t)=\frac{1}{t}\int_0^tdu \;_3F_2(5/4,3/2,7/4;2,2;u).\eqno(A2)$$
However,
$$\;_3F_2(5/4,3/2,7/4;2,2;u)=\frac{32}{3}\frac{d}{du}\;_3F_2(1/4,1/2,3/4;1,1;u).\eqno(A3)$$
Since,
$$\;_3F_2(a,b,a+1/2;1,2a+b;z)$$
$$=\left[\frac{2}{z}(1-\sqrt{1-z})\right]^{2a}\;_3F_2(2a,2a,1-b;1,2a+b;1-\frac{2}{z}(1-\sqrt{1-z})),\eqno(A4)$$
by combining these relations, after integrating by parts and noting that
$$\;_3F_2(1/2,1/2,1/2;1,1;z)=\frac{4}{\pi^2}{\bf K}^2\left(\sqrt{\frac{1-\sqrt{1-z}}{2}}\right),\eqno(A5)$$
 after  change of  integration variable, one finds
$$\;_5F_4(1,1,5/4,3/2,7/4;2,2,2,2;z)=$$
$$\frac{256}{3\pi^2z}\int_0^{\alpha(z)}\frac{dt}{t}\frac{(1-6t^2+t^4)}{1-t^4}\left[(1+t^2){\bf K}^2(t)-{\bf K}^2(0)\right]\eqno(A6)$$
$$\alpha(z)=\sqrt{\frac{\sqrt{2}-\sqrt{1+\sqrt{1-z}}}{\sqrt{2}+\sqrt{1+\sqrt{1-z}}}}.$$
This shows analyticity in the $z$-plane cut along the positive real axis for $z\ge1$.


\section*{References} 

\noindent
[1] C. Deninger, J. Amer. Math. Soc.{\bf 10},259-281 (1997)

\noindent
[2] D.W. Boyd, Experimental Math. {\bf 7}, 37-82 (1998)

\noindent
[3] M.J. Bertin, J. Num. Theory {\bf 128}, 2890-2913 (2008)

\noindent
[4] M. Rogers, Ramanujan J. {\bf 18},327-340(2011)

\noindent
[5] M. Rogers, IMRN {\bf 17},4027-4058 (2011)

\noindent
[6] F.Y. Wu, J.Phys. A:Math. Gen.{\bf 10}, L113-L115(1977)

\noindent
[7] R. Shrock and F.Y. Wu, J. Phys. A:Math.Gen.{\bf 38},3881-3902(2007)

\noindent
[8] F.Y. Wu and M.L. Glasser, Ramanujan J. {\bf10}, 205-215(2005)

\noindent
[9] M.L. Glasser and G. Lamb, J. Phys. A:Math. Gen.{38}, L471-L473(2005)

\noindent
[10]  I.J. Zucker, J. Stat. Phys {\bf145}, 591-612(2011) 

\noindent
[11] A. Rosegren, J. Phys. A:Math.Gen.{\bf20}, L993-L927 (1987)

\noindent
[12] G.S. Joyce and I.J. Zucker, J.Phys. A:Math.Gen.{\bf34}, 7349-7354 (2001)

\noindent
[13]  M.L. Glasser and E. Montaldi, Physical Rev. {\bf E48}, R2340-R2342 (1993)

\noindent
[14] G.S. Joyce, Phil. Trans. Roy. Soc. London {\bf A273}, 573-610(1973)

\noindent
[15] G.S. Joyce, J. Phys. A: Math. Theor.{\bf 45}, 285001(2012).

\noindent
[16] A.J. Guttmann and M. Rogers, Arxiv:math-phys.1207.2815v1 (2012); 

\noindent
[17]  M.L. Glasser and A.J. Guttmann, J. Phys. A:Math. Gen.{\bf 27}, 7011-7014 (1974)

\end{document}